\RequirePackage{ifpdf}
\documentclass[12pt,aps,amsmath,amsfonts]{article}
\usepackage{jheppub}

\include{mydefsFractionalLapl}
\usepackage{changes,amsmath,mathtools}
\usepackage{graphicx}
\usepackage{epsfig}
\usepackage{amsfonts,amssymb,amsmath,bm}
\usepackage{enumitem}
\usepackage[hang]{subfigure}
\usepackage{epstopdf}
\usepackage{mathtools}
\usepackage{color}
\usepackage{mathrsfs}
\usepackage{esint}

\usepackage[overload]{empheq}
\usepackage{cases}
\newcommand{\beq}{\begin{equation}}
\newcommand{\eeq}{\end{equation}}
\newcommand{\ba}{\begin{eqnarray}}
\newcommand{\ea}{\end{eqnarray}}

\definecolor{airforceblue}{rgb}{0.36, 0.54, 0.66}
\definecolor{steelblue}{rgb}{0.27, 0.51, 0.71}
\definecolor{amber}{rgb}{1.0, 0.49, 0.0}

\title{
Quantization of nonlocal fractional field theories via the extension problem
}

\author[a]{Antonia M. Frassino}
\author[b]{Orlando Panella}

\affiliation[a]{Departament de F{\'\i}sica Qu\`antica i Astrof\'{\i}sica, Institut de
Ci\`encies del Cosmos,\\ Universitat de
Barcelona, Mart\'{\i} i Franqu\`es 1, E-08028 Barcelona, Spain}
\affiliation[b]{ INFN, Sezione di Perugia, Via A Pascoli, I-06123, Perugia, Italy }

\emailAdd{antoniam.frassino@icc.ub.edu}
\emailAdd{orlando.panella@cern.ch}

\abstract{
We use the extension problem proposed by Caffarelli and Silvestre to study the quantization of a scalar nonlocal quantum field theory  built out of the fractional Laplacian. We show that the quantum behavior of such a nonlocal field theory in $d$-dimensions can be described in terms of a local action in $d+1$ dimensions  which can be quantized using the canonical operator formalism though giving up local commutativity. In particular, we  discuss how to obtain the two-point correlation functions and the vacuum energy density of the nonlocal fractional theory as a brane limit of the bulk correlators. We show explicitly how the quantized extension problem reproduces exactly the same particle content of other approaches based on the spectral representation of the fractional propagator. We also briefly discuss the inverse fractional Laplacian and possible applications of this approach in general relativity and cosmology. 
}

\keywords{Nonlocal theory, fractional Laplacian, Extra dimensions, Vacuum energy. }
\date{} 

\begin{document}

\maketitle

\section{Introduction}
During the past years, there has been an increasing interest in the study of nonlinear and quasi-linear equations that involve fractional powers of the Laplacian (see for example \cite{bjorland2012non,barrios2018global,silvestre2007regularity, doi:10.1002/cpa.21379}). 
In particular, fractional powers of the Laplacian are considered in the case of nonlocal diffusion processes described  by nonlinear partial differential equations. In such processes, differently from what happens in the standard case for the Heat equation, the interactions between particles could be nonlocal, which requires  the use of more general operators than the standard Laplacian. 
The study of nonlinear partial differential equations is complicated also by the fact that there is not a specific approach to investigate the properties of the solutions (e.g., existence, uniqueness, regularity, asymptotic
behavior, velocity of propagation, or feasible numerical methods).

In contrast with the usual Laplacian operator, 
the fractional Laplacian is a nonlocal operator in the sense that its value at any point depends on the value of the function everywhere else. For  $\alpha \in (0,2)$ the fractional Laplacian $(-\Delta)^{\frac{\alpha}{2}}$ of a function $f:\mathbb{R}^n \rightarrow \mathbb{R}$ can be defined through the Fourier transform as
\begin{equation}\label{DeltaFourier}
(-\Delta)^{\frac{\alpha}{2}} f(x)=\int \frac{d^nk}{(2\pi)^n}\, |k|^{\alpha}\tilde{f}(k) e^{ikx}\, ,
\end{equation}
that for sufficiently smooth functions can be rewritten as a singular integral operator as
\begin{equation}
\left( - \Delta \right)^{\frac{\alpha}{2}} \;f(x) = C_{n,\alpha}\int_{\mathbb{R}^n} \frac{f(x) - f(x')}{|x-x'|^{n+\alpha}} d^nx'
\end{equation}
where the real parameter  $C_{n,\alpha}=2^{\alpha}\Gamma (\frac{n+\alpha}{2})/[ \pi^{n/2}|\Gamma(-\alpha/2)|]$ is a normalization constant. This definition shows that the fractional Laplacian is a nonlocal operator.
However,
there are also other equivalent possible definitions of the fractional Laplacian \cite{Mateusz:2017aa} that will turn out to be useful in the following discussion.
Interestingly, given a function $f:\mathbb{R}^n \rightarrow \mathbb{R}$, the following extension problem 
 \begin{subnumcases}{}
    \!\! u(x,0)=f(x) \qquad x \in \mathbb{R}^n \label{eq:bc} \\
    \!\! \nabla \cdot \left( y^{1-\alpha} \nabla u \right) = 0 \qquad x \in \mathbb{R}^n, y>0\, , \label{eq:eom}
\end{subnumcases}
can be solved to obtain a smooth bounded function $u:\mathbb{R}^n \times [0,\infty) \rightarrow \mathbb{R}$, where
the nonlocal operator $(- \Delta)^{\frac{\alpha}{2}}$ satisfies \cite{Caffarelli:2007aa} 
\begin{equation} \label{eq:Delta}
(- \Delta)^{\frac{\alpha}{2}} f(x) =\, +\, C\lim_{y\rightarrow 0^+} y^{1-\alpha}u_y (x,y).
\end{equation}
This is a Dirichlet-to-Neumann operator for an appropriate
harmonic extension problem and $C$ is a constant that allows to accurately define the limit on the RHS of \eqref{eq:Delta}. The constant $C$ is defined as \cite{CHANG20111410} 
\begin{equation} \label{eq:preConst}
    C = \frac{2^{\alpha} \Gamma \left[\alpha / 2 \right]}{ \alpha \, \Gamma \left[-\alpha/2 \right]}.
\end{equation}
The notable property presented by Eq. \eqref{eq:Delta} is that the nonlocal operator $(-\Delta)^{\frac{\alpha}{2}}$ acting on $f(x)$ in the domain of $\Delta$ is localized through $u$.
It has been shown in \cite{CHANG20111410} that the problem can also be extended to the wider range $\alpha \in [0,n]$ 
(and in \cite{doi:10.1080/03605301003735680} that the same extension problem is also valid for negative values of the constant $\alpha$).
Furthermore, the characterization  given in Eq. \eqref{eq:Delta} has also
been used to show that the fractional Laplacian $(-\Delta)^{\frac{\alpha}{2}}$ coincides with a certain conformally covariant
operator $P_{\frac{\alpha}{2}}$, on the hyperbolic space $\mathbb{R}^{n+1}_+$ from which $\mathbb{R}^n$ is to be seen as its boundary \cite{CHANG20111410}. 
\\
In general, the fractional Laplacian has been used in different fields in physics, for example in solid mechanics to model the elastic behavior of nonlocal continua  \cite{SAPORA201363,PAOLA20085642}, in quantum mechanics and optics \cite{2015OptL...40.1117L}
or in the context of holography, in models with nonlocality
due to charge screening \cite{Limtragool:2016gnl, LaNave:2017nex}.
Here, we will also discuss connections to the field of cosmology, and in particular to some modified theories of gravity introduced in the past years that either make use of extra dimensions or of the presence of nonlocal terms in the Lagrangian so to model
the current phase of cosmological acceleration without resorting to a cosmological constant \cite{Maggiore:2013mea,Maggiore:2014sia,Belgacem:2017cqo,ArkaniHamed:2002fu,Barvinsky:2003kg,Deser:2007jk}. 
Interestingly, an analysis of the connection between nonlocal field theories in $d$-dimensions and local field theories in $(d+1)$-dimensions has been explored from the Hamiltonian point of view in \cite{Gomis:2000gy}.

In this work by using the extension problem it will be shown that quantizing a local field theory in a  bounded space-time (bulk) with an extra space-like transverse dimension ($y \in [0,+\infty ]$) via the operator formalism, but waiving local commutativity,  reproduces,  through a limiting procedure ($y\to 0$), the correlation functions of the non-local theory showing a complete equivalence with the fractional propagator obtained independently via analiticity considerations in terms of the spectral representation.

The structure of the paper is the following: In Sec.~\ref{sec:Action} we analyze the connection between the bulk local action and the nonlocal action on the brane.
In Sec.~\ref{sec:quantization} we present the quantization of the local action in $(4+1)$-dimensions via the operator formalism. We will see that the two-point bulk correlation functions induce, on the brane, the expected nonlocal correlations functions.
As mentioned before, the extension problem has also been shown to be useful in the case of nonlocal terms in the form of the inverse Laplacian. We will study this point in Sec.~\ref{sec:negative-power}. A final discussion about other fields is reported in Sec.~\ref{sec:discussion}.

\section{Nonlocal action properties \label{sec:Action}}

Although we have in mind physical applications of the Caffarelli-Silvestre extension problem to a fractional scalar field theory in Minkowski space-time, we stick for the time being to the mathematicians notation and use the $\Delta$ operator. Ordinary results applicable to Minkowski space-time  can be recovered after a Wick rotation of the $x_0$ component. 

Let us, therefore, consider a nonlocal scalar theory in $d$ Euclidean dimensions.
Using the extension problem \eqref{eq:eom}, the following equality holds:
\begin{equation} \label{eq:nonlocTOloc}
    S= -\,\frac{1}{2\,C}\, \int d^d x\, \phi (x) \left(-\Delta \right)^{\frac{\alpha}{2}} \phi (x) =\frac{1}{2}\,
     \int d^{d}x \int^{\infty}_{0} dy\;y^{1-\alpha} \partial_{\mu} \Phi (x,y) \partial^{\mu} \Phi \left( x,y \right)
\end{equation}
where $C$ is the constant \eqref{eq:preConst}, $\Phi (x,y)$ denotes the scalar field in $(d+1)$ dimensions and $ \partial_{\mu}$ denotes the  derivatives with respect to the $d+1$ variables $(\{x\},y)$.
To prove the equivalence \eqref{eq:nonlocTOloc} let us start with the following action in $(d+1)$ Euclidean dimensions:
\begin{eqnarray}
S= \frac{1}{2}\,\int_{-\infty}^{+\infty} dx  \int_{y>0}  \;dy \;y^{1-\alpha} \left[ \partial_{x} \Phi \; \partial_{x} \Phi + \partial_{y} \Phi \; \partial_{y} \Phi \right]
\end{eqnarray}
then, integration by parts gives
\begin{eqnarray}
S &=& \frac{1}{2}\,\int_{0}^{\infty} dy\; y^{1-\alpha} 
\left[  \Phi \partial_x \Phi \bigg|_{-\infty}^{+\infty} - \int_{-\infty}^{\infty} dx \; \Phi \Delta_{x} \Phi \right] + \nonumber \\
&+ & \frac{1}{2}\, \int_{-\infty}^{\infty} dx \, \Phi\, 
\left[  y^{1-\alpha} \Phi \partial_y \Phi \bigg|_{0}^{+\infty} - \int_{0}^{+\infty}dy \left[ \left( 1-\alpha \right) y^{-\alpha} \partial_{y} \Phi +y^{1-\alpha} \Delta_{y} \Phi \right] 
\right].
\end{eqnarray}
Now, supposing that the field $\Phi(x,y)$ at $x^i=\pm \infty$ and $y=+\infty$ is zero, we are left only with
\begin{eqnarray}
\label{eq:Sbyparts}
S &=& - \frac{1}{2}\,\int_{-\infty}^{\infty} dx  \int_{0}^{+\infty}dy \; 
\Phi \left[
 \Delta_{x} \Phi + \left( 1-\alpha \right) y^{-\alpha} \partial_{y} \Phi +y^{1-\alpha} \Delta_{y} \Phi
\right]
+ \nonumber\\
 && - \frac{1}{2}\,\int_{-\infty}^{\infty} dx \lim_{y\rightarrow 0^{+}} \left[ y^{1-\alpha} \Phi \; \partial_y \Phi \right]
\end{eqnarray}
and using Eq.~\eqref{eq:bc} and  Eq.~\eqref{eq:eom} 
\begin{eqnarray}
\text{div} \left( y^{1-\alpha} \nabla u \right) 
&=& y^{1-\alpha} u_{xx} + y^{1-\alpha} u_{yy}  + (1-\alpha)y^{-\alpha} u_y = 0,
\end{eqnarray}
in the first line of Eq.~\eqref{eq:Sbyparts} and
then, substituting Eq.~\eqref{eq:Delta} in the second line, the action \eqref{eq:Sbyparts} reduces to 
\begin{eqnarray}
S= -\, \frac{1}{2 \, C} \int_{-\infty}^{\infty} dx \; \phi(x) \left( - \Delta \right)^{\frac{\alpha}{2}} \phi(x).
\end{eqnarray}
Hence, using the extension problem, a nonlocal theory can be described in terms of a local action in a $(d+1)$ spacetime as expressed in Eq.~\eqref{eq:nonlocTOloc}. 
The equivalence~\eqref{eq:nonlocTOloc}
relates, therefore, a local quantum field theory in the bulk (hyperbolic) space $\mathbb{R}^{n+1}_+$ to a  nonlocal quantum field theory on the non-trivial boundary $\mathbb{R}^{n}$~\cite{Rajabpour:2011qr}.
\\
It should be noted indeed that the coefficient present in the $(d+1)$ dimensional part of the Eq.~\eqref{eq:nonlocTOloc}, can be associated  to  the  half space ($y>0$) representation  of an Anti-de Sitter spacetime (or hyperbolic space if Wick rotated):
\begin{equation}
    ds^2 = \frac{1}{y^2} \left( - dt^2 + dy^2 + \sum_{i=2}^{d} dx_{i}^2  \right) = \left( \frac{1}{y^2} \eta_{\mu \nu} \right) dx^{\mu} dx^{\nu} = g_{\mu \nu} dx^{\mu} dx^{\nu}.
\end{equation}
Therefore, starting from the RHS of Eq.~\eqref{eq:nonlocTOloc} follows
\begin{equation}
   y^{1-\alpha} \eta^{\mu \nu} \partial_{\mu} \Phi \partial_{\nu} \Phi = y^{1-\alpha-2} y^{2}\eta^{\mu \nu} \partial_{\mu} \Phi \partial_{\nu} \Phi = \sqrt{\det g}\;\;   g^{\mu \nu} \partial_{\mu} \Phi \partial_{\nu} \Phi
\end{equation}
where the last equality is valid only if $\alpha = d$.

\subsection{Variation of the nonlocal action}
Here let us consider the nonlocal action by itself and derive the corresponding e.o.m. by computing the variation of the action:
\begin{equation}
\label{NLaction}
S=\frac{1}{2\, C}\, \int \, d^4x\,\phi(x) (-\Delta)^\frac{\alpha}{2}\, \phi(x)
\end{equation}
under an arbitrary variation of the field $\delta\phi(x)$:
\begin{equation}
\phi(x) \to \phi(x) +\delta\phi(x)
\end{equation}
The variation of the action to linear order in $\delta\phi(x)$ is given by:
\begin{eqnarray}
\label{var1}
 \delta S &=& S[\phi +\delta\phi] -S[\phi] \nonumber   \\
 &=& \frac{1}{2\, C}\, \int \, d^4x\,\left( \phi(x) +\delta\phi(x)\right) (-\Delta)^\frac{\alpha}{2}\, \left(\phi(x) +\delta\phi(x)\right) -\frac{1}{2\, C}\, \int \, d^4x\,\phi(x) (-\Delta)^\frac{\alpha}{2}\, \phi(x)\nonumber\\&=&
 \frac{1}{2\,C}\, \int \, d^4x\, \phi(x) (-\Delta)^\frac{\alpha}{2}\, \delta\phi(x)  +\frac{1}{2\, C}\, \int \, d^4x\,\delta\phi(x) (-\Delta)^\frac{\alpha}{2}\, \phi(x) +{\cal O}(\delta\phi^2)
\end{eqnarray}
Now we use the distributional definition of the fractional Laplacian stated in (b) of theorem 1.1 of~\cite{Mateusz:2017aa}:
\begin{equation}
 \int \, d^4x\, f(x) (-\Delta)^\frac{\alpha}{2}\, g(x)= \int \, d^4x\, g(x) (-\Delta)^\frac{\alpha}{2}\, f(x)  \, ,
\end{equation}
in order to show that  the first term on the RHS of Eq.~\eqref{var1} is the same  as  the second (with the Laplacian acting on the field $\phi(x)$). So Eq.~\eqref{var1} becomes:
\begin{equation}
  \label{var2}  
  \delta S = \frac{1}{C}\,\int \, d^4x\,\delta\phi(x) (-\Delta)^\frac{\alpha}{2}\, \phi(x) +{\cal O}(\delta\phi^2)
\end{equation}
and the e.o.m. are obtained requiring that the first order (in $\delta\phi$) variation of the action vanishes. Since the variation is arbitrary we obtain the e.o.m. as:
\begin{equation}
    (-\Delta)^\frac{\alpha}{2} \phi(x) =0
    \label{fraclaplaceeq}
\end{equation}
i.e. the fractional Laplace equation  Eq.~\eqref{fraclaplaceeq} sets an important condition: it corresponds to the particular case in which the RHS of Eq.~\eqref{eq:Delta} is equal to zero.
The e.o.m., therefore, sets the following boundary condition for the quantum field $\Phi(x,y)$ propagating in the bulk space with the extra dimension $y$: 
\begin{equation}
    \lim_{y\downarrow 0} y^{1-\alpha} \Phi_{y} (x,y )=0.
\end{equation}
This boundary condition will play a crucial role in the quantization of the theory.
Some aspects of boundary conditions in the bulk and their effect on the field dynamics on the brane  have been analyzed in \cite{Mintchev:2001mh,Mintchev:2001aa}.
In particular in \cite{Mintchev:2001aa} it is shown how a large family of bulk fields quantized giving up local commutativity induces brane fields (in the limit $y\to 0$) which  do verify it.

\section{Quantization of the extended problem in $(d+1)$ dimensions\label{sec:quantization}}
In this section we perform a quantization of  the local action in $(4+1)$-dimensions  via the operator formalism and show that its two-point bulk correlation functions induce, on the brane, the expected nonlocal correlations functions as in Eq.~\eqref{fracprop}. 
\subsection{Eigenfunctions}
 We now switch to Minkowski space-time performing a Wick rotation $x_0=ict$ in Eq.~\eqref{eq:nonlocTOloc} and obtain: 
\begin{equation} \label{eq:nonlocTOloc2}
    S= \,\frac{1}{2\,C}\, \int d^d x \phi (x) \Box^{\frac{\alpha}{2}} \phi (x) = \,\frac{1}{2}\, 
     \int d^{d}x \int^{\infty}_{0} dy\;y^{1-\alpha} \partial_{\mu} \Phi (x,y) \partial^{\mu} \Phi \left( x,y \right)\, .
\end{equation}

\noindent Let us therefore start with the following action for a massless scalar field $\Phi(x,y)$ in the bulk
\begin{equation}
    S=\frac{1}{2}\int d^{d}x \int^{\infty}_{0} dy\;y^{1-\alpha} \partial_{\mu} \Phi (x,y) \partial^{\mu} \Phi \left( x,y \right), 
    \label{LocalAction}
\end{equation}
the variation of the action gives the equation of motion $\partial^{\mu} \left( y^{1-\alpha} \partial_{\mu} \Phi(x,y) \right)=0$, that explicitly reads
\begin{equation}
    \label{eq:eqom}
   \Box_{x} \Phi (x,y) - (1-\alpha) \frac{1}{y} \partial_{y} \Phi (x,y) - \partial^{2}_{y} \Phi (x,y)= 0,
\end{equation}
where $\Box_{x}=\partial_{0}^{2}-\nabla^{2}$.  Eq. \eqref{eq:eqom} can be solved by separation of  variables, writing $\Phi (x,y) = \varphi (x) \psi(y)$, from which follows
\begin{equation}
   \frac{1}{\varphi(x)} \Box_{x}\varphi (x) = \frac{1}{\psi(y)} \left[ \partial^{2}_{y} + (1-\alpha)\frac{1}{y}\partial_{y} \right] \psi(y) = \text{const} = -\lambda^2,
\end{equation}
where $\lambda$ has dimension of mass. Therefore, from the e.o.m., one gets the following equations:
\begin{subequations}
\label{eq:eom_separated}
\begin{align}
 & \left( \Box_{x} + \lambda^2 \right) \varphi(x) = 0, \label{eq:eom_separateda}\\
 &  \left[ \partial^{2}_{y} + (1-\alpha)\frac{1}{y} \partial_y \right] \psi(y) = - \lambda^2 \psi(y). \label{eq:eom_separatedb}
\end{align}
\end{subequations}
The first one, Eq.~\eqref{eq:eom_separateda}, gives the usual expansion of the field $\varphi(x)$ in plane waves of mass $\lambda$, whereas the second, Eq.~\eqref{eq:eom_separatedb}, can be solved imposing the appropriate boundary condition:
\begin{equation}
   y^{1-\alpha}\partial_{y} \psi(y)=0.
\end{equation}
This boundary condition on the $y$-dependent part of the wave-function comes from the condition in Eq.~\eqref{eq:Delta} on the field $\Phi(x,y)$.
The equation for the wave-function $\psi(y)$ can be reduced to a Schr\"{o}dinger problem with a potential of the type $1/y^2$ (see \cite{Presilla:2015qda}). 
Therefore, in order to find the eigen-functions $\psi(y)$, one can proceed in the following way. The first step is to change the variable  $t=\lambda y$ in Eq.~\eqref{eq:eom_separatedb}
\begin{equation}
   \left[ \lambda^2 \partial^{2}_{t} + (1-\alpha) \frac{\lambda}{t} \; \lambda \partial_t \right] \psi = -\lambda^2 \psi \, .
\end{equation}
Then, writing $\psi(t) = t^{\frac{\alpha}{2}} \chi(t)$,  one gets the  Bessel equation 
\begin{equation}
   t^2 \partial^{2}_{t} \chi(t) + t \partial_{t} \chi(t) + \left[ t^2 - \left( \frac{\alpha}{2} \right)^2 \right] \chi(t) =0,
\end{equation}
whose general solution  is the linear combination of $J_{\alpha/2}(t)$ and $Y_{\alpha/2}(t)$: 
\begin{equation}
   \psi(\lambda,y) = t^{\frac{\alpha}{2}} \left[ A J_{\alpha/2}(t) + B Y_{\alpha/2}(t) \right]_{t=\lambda y}. \label{eq:psiBessel}
\end{equation}
The two constants $A$ and $B$ in Eq.~\eqref{eq:psiBessel} can be fixed imposing the boundary conditions 
\begin{equation} \label{eq:bcy}
   \lim_{y\rightarrow 0} \psi(\lambda,y)=1, \qquad \lim_{y \rightarrow 0} y^{1-\alpha} \partial_{y} \psi(\lambda,y)=0. 
\end{equation}
Using the asymptotic expansion for $t\rightarrow 0$ (i.e., $y\rightarrow 0$)
\begin{eqnarray}
 J_{\alpha/2} (t) & \simeq & \frac{t^{\frac{\alpha}{2}}}{ 2^{\frac{\alpha}{2}} \; \Gamma\left( \frac{\alpha}{2} +1 \right)}\, , \\
 Y_{\alpha/2} (t) & \simeq & 
 -  \frac{2^{\frac{\alpha}{2}} \Gamma \left(\frac{\alpha }{2}\right) t^{-\frac{\alpha }{2}}}{\pi }
 + \frac{\cot\left( \frac{\alpha}{2} \pi \right) t^{\frac{\alpha}{2}}}{\Gamma \left( \frac{\alpha}{2}+1 \right)\;2^{\frac{\alpha}{2}} } \, ,
\end{eqnarray}
the solution \eqref{eq:psiBessel} reads
\begin{equation}
   \psi(\lambda, y) \simeq \frac{ \left[ A + B\cot\left( \pi \frac{\alpha}{2} \right) \right] t^{\alpha}}{ \Gamma \left( \frac{\alpha}{2} +1 \right) 2^{\frac{\alpha}{2}}} - \frac{B\; \Gamma \left( \frac{\alpha}{2} \right) 2^{\frac{\alpha}{2}}}{\pi}\, .
   \label{eq:psiBessellty0}
\end{equation}
Then, the first boundary condition in \eqref{eq:bcy} gives 
\begin{equation}
   \lim_{y\rightarrow 0} \psi(\lambda, y) = -\frac{B\; \Gamma\left( \frac{\alpha}{2} \right) 2^{\frac{\alpha}{2}}}{\pi} = 1\, ,
\end{equation}
and the second boundary condition in Eq.~\eqref{eq:bcy} gives
\begin{equation}
   \lim_{y \rightarrow 0} y^{1-\alpha} \partial_{y} \psi(\lambda,y)=0 \Leftrightarrow A+B\; \cot\left( \pi \frac{\alpha}{2} \right) = 0.
\end{equation}
Thus, the eigenfunctions \eqref{eq:psiBessel} read
\begin{equation}
   \psi(\lambda,y)=\frac{\pi \left(\lambda y \right)^{\frac{\alpha}{2}}}{\Gamma\left( \frac{\alpha}{2} \right) 2^{\frac{\alpha}{2}}} 
   \left[ \cot\left( \pi \frac{\alpha}{2} \right) J_{\alpha/2}(\lambda y) - Y_{\alpha/2}(\lambda y)
   \right]. \label{eq:psiBessel2}
\end{equation}
Now, using the definition
\begin{equation}
   Y_{\nu}(z) = J_{\nu} \cot(\pi \nu) - \frac{1}{\sin (\pi \nu)} J_{-\nu}(z)\, ,
\end{equation}
follows the identity
\begin{equation}
   J_{\alpha/2}(\lambda y) \cot (\pi \frac{\alpha}{2}) - Y_{\alpha/2}(\lambda y) = \frac{1}{\sin (\pi \frac{\alpha}{2})} J_{-\alpha/2}(\lambda y)
\end{equation}
that, used in \eqref{eq:psiBessel2}, gives the final form
\begin{equation}
   \psi(\lambda , y) = \frac{\pi}{2^{\frac{\alpha}{2}} \Gamma (\frac{\alpha}{2}) \sin \left(\pi \frac{\alpha}{2} \right)} (\lambda y)^{\frac{\alpha}{2}} J_{-\alpha/2} (\lambda y). \label{eq:psiBessel3}
\end{equation}
This result can be compared with the final result of  Sect.~3.2 of \cite{doi:10.1080/03605301003735680}, where the authors study the existence and uniqueness results for the extension problem.
Note that the eigenfunctions \eqref{eq:psiBessel3} satisfy the boundary conditions
\begin{equation}
   \psi(\lambda, 0) =1,\qquad y^{1-\alpha} \partial_{y} \left. \psi(\lambda, y) \right|_{y=0} =0. 
\end{equation}
as required in \eqref{eq:bcy}.
These eigenfunctions form an orthonormal system and satisfy the completeness relation
\begin{eqnarray}
    \label{eq:complRel}
   \int d\lambda \; \mu(\lambda) \bar{\psi}(\lambda, y_{1}) \psi(\lambda, y_2) 
   &=&  
   \frac{ \left( 2^{-\frac{\alpha}{2}} \pi \right)^{2} y_{1}^{\frac{\alpha}{2}} y_{2}^{\frac{\alpha}{2}}}{ \left[ \Gamma (\frac{\alpha}{2}) \sin \left( \frac{\pi \alpha}{2} \right) \right]^2} 
   \intop_{0}^{\infty} d\lambda \lambda^{\alpha} \mu(\lambda) J_{-\frac{\alpha}{2}} (\lambda y_{1}) J_{-\frac{\alpha}{2}} (\lambda y_{2}).\qquad
\end{eqnarray}
For the Bessel functions, the following identity~\cite{NIST:DLMF} holds\footnote{In relation to the notation in subsection \ref{subsec:unparticle}, this means that the identity holds for $d_{\cal U}>1$.} for  $\nu > -1$ :
\begin{equation}
   \intop_{0}^{\infty} d\lambda\;\lambda\; J_{\nu}(\lambda y_{1}) J_{\nu}(\lambda y_{2}) = \frac{1}{y_{1}} \delta \left( y_{1} - y_{2} \right)\, .
\end{equation}
We now consider a function $\mu (\lambda)$ in Eq.~\eqref{eq:complRel} such that
\begin{equation}
   \left[  \frac{2^{-\frac{\alpha}{2}} \pi}{ \Gamma (\frac{\alpha}{2}) \sin \left(\pi \frac{\alpha}{2} \right)} \right]^{2} \mu(\lambda)\; \lambda^{\alpha} = \lambda
\end{equation}
so that:
\begin{equation}
\label{mufuntion}
   \mu (\lambda)= \left[  \frac{2^{-\frac{\alpha}{2}} \pi}{ \Gamma (\frac{\alpha}{2}) \sin \left(\pi \frac{\alpha}{2} \right)} \right]^{-2}\; \lambda^{1-\alpha}\, .
\end{equation}
Then the eigenfunctions satisfy the completeness relation:
\begin{equation}
\label{completeness}
   \intop_{0}^{\infty}\;d\lambda\, \mu (\lambda) \bar{\psi} (\lambda, y_{1}) \psi (\lambda, y_{2}) = \frac{(y_1y_2)^{\frac{\alpha}{2}}}{y_1} \delta(y_{1} - y_{2}) \,\, =\,\, y_1^{\alpha-1}\, \delta(y_{1} - y_{2}) \,.
\end{equation}
This particular choice of the function $\mu(\lambda)$ is due to our interest in describing conformal (scale) invariance on the brane. As it will be clear in the following section the function $\mu(\lambda)$  plays a central role in the quantization of the theory. It will ensure the canonical  equal-time commutation relation between the bulk field and its conjugate momentum.
As discussed in \cite{Mintchev:2001aa} in general other  possibilities could be considered for the function $\mu(\lambda)$ beyond the one in Eq.~(\ref{mufuntion}) but requiring scale invariance fixes it to the specific form $\mu(\lambda)\propto \lambda^{1-\alpha}$. This will be discussed again in connections with massive models of gravity in Section \ref{sec:discussion}.

\subsection{Quantization rules and field expansion}
The field decomposition over its normal modes is:
\begin{eqnarray} \label{eq:def-Phi}
    {\Phi}(x,y) &=& \int \frac{d^3 \bm{p}}{(2\pi)^3}\, \intop_0^{+\infty} d\lambda\, \mu(\lambda) \left[ a(\bm{p},\lambda) \frac{e^{i[\bm{p}\cdot\bm{x}-\omega_{\lambda^2}(\bm{p})t]}}{\sqrt{2\,\omega_{\lambda^2}(\bm{p})}} \psi(y,\lambda) \right. \nonumber \\&&\phantom{xxxxxxxxxxxxxxxx}\left. +a^\dagger(\bm{p},\lambda) \frac{e^{-i[\bm{p}\cdot\bm{x} -\omega_{\lambda^2}(\bm{p})t]}}{\sqrt{2\,\omega_{\lambda^2}(\bm{p})}} \psi^*(y,\lambda)\right] 
\end{eqnarray}
where $\omega_{\lambda^2}(\bm{p})=\sqrt{\bm{p}^2+\lambda^2}$,
and $a(\bm{p},\lambda),a^\dagger(\bm{p},\lambda)$ are creation and annihilation operators for the modes of momentum $\bm{p}$ and mass $\lambda$ with quantization rules~\cite{Mintchev:2001aa}:
\begin{subequations}
\label{quantization-rules1}
\begin{align}
\label{quantization-rules1a}
&\left[a(\bm{p},\lambda),a^\dagger(\bm{p},\lambda)\right] 
=(2\pi)^3\delta^3(\bm{p}-\bm{p}') \frac{1}{\mu(\lambda)}\delta(\lambda-\lambda')\\
&\left[a(\bm{p},\lambda),a(\bm{p},\lambda)\right]=0\label{quantization-rules1b}\\
&\left[a^\dagger(\bm{p},\lambda),a^\dagger(\bm{p},\lambda)\right]=0\label{quantization-rules1c}
\end{align}
\end{subequations}
To find the equal-time commutation relations we
start from Eq.~\eqref{eq:def-Phi} and take the derivative of the field:
\begin{eqnarray}
    \partial_{0}  {\Phi}(x,y) &=&\int \frac{d^3 \bm{p}}{(2\pi)^3}\, \intop_0^{+\infty} d\lambda\, \mu(\lambda) \left[ a(\bm{p},\lambda) \frac{e^{i[\bm{p}\cdot\bm{x}-\omega_{\lambda^2}(\bm{p})t]}}{\sqrt{2\,\omega_{\lambda^2}(\bm{p})}} 
    \left[ -i \omega_{\lambda^2}(\bm{p}) \right]
    \psi(y,\lambda) \right. \nonumber \\&&\phantom{xxxxxxxxxxxxx}\left. +a^\dagger(\bm{p},\lambda) \frac{e^{-i[\bm{p}\cdot\bm{x} -\omega_{\lambda^2}(\bm{p})t]}}{\sqrt{2\,\omega_{\lambda^2}(\bm{p})}}
    \left[ i \omega_{\lambda^2}(\bm{p}) \right]
    \psi^*(y,\lambda)\right].
\end{eqnarray}
The conjugate momentum of $\Phi$, derived from the Lagrangian density of the local action in Eq.~\eqref{LocalAction}, is
\begin{equation}
\label{conjugate_momentum_def} 
\Pi(t,\bm{r},y)= y^{1-\alpha}\,\partial_t \Phi(t,\bm{r},y)\, .
\end{equation}
We then compute
\begin{eqnarray}
    &&\left[ \partial_{0}  {\Phi}(\bm{r},y,t) ,{\Phi}(\bm{r}^{\prime},y^{\prime},t)  \right]
    =  \frac{1}{2}  \int \frac{d^3 \bm{p}}{(2\pi)^3}\, \int \frac{d^3 \bm{p}^{\prime}}{(2\pi)^3}\,
    \intop_0^{+\infty} d\lambda\, \mu(\lambda)  \intop_0^{+\infty} d\lambda^{\prime}\, \mu(\lambda^{\prime}) \times \\
    &&
    \left[ a(\bm{p},\lambda)\, \psi(y,\lambda) \, e^{i[\bm{p}\cdot\bm{r}-\omega_{\lambda^2}(\bm{p})t]}
    (-i) \sqrt{  \omega_{\lambda^2}(\bm{p}) }
     +a^\dagger(\bm{p},\lambda) \,\psi^*(y,\lambda)
     e^{-i[\bm{p}\cdot\bm{r} -\omega_{\lambda^2}(\bm{p})t]}
     i \sqrt{ \omega_{\lambda^2}(\bm{p})} , \right. \nonumber\\
    &&
    \left.
    a(\bm{p}^{\prime},\lambda^{\prime})\, \psi(y^{\prime},\lambda^{\prime}) \, \frac{e^{i[\bm{p}^{\prime}\cdot\bm{r}^{\prime}-\omega_{{\lambda^{\prime}}^2}(\bm{p}^{\prime})t]}}{ \sqrt{  \omega_{{\lambda^{\prime}}^2}(\bm{p}^{\prime}) }}
     +a^\dagger(\bm{p}^{\prime},\lambda^{\prime}) \,\psi^*(y^{\prime},\lambda^{\prime})
     \frac{e^{-i[\bm{p}^{\prime}\cdot\bm{r}^{\prime} -\omega_{{\lambda^{\prime}}^2}(\bm{p}^{\prime})t]}}
     { \sqrt{ \omega_{{\lambda^{\prime}}^2}(\bm{p}^{\prime})}}
    \right]\, . \nonumber
\end{eqnarray}
Applying the quantization rules \eqref{quantization-rules1a}-\eqref{quantization-rules1c} only two of the four terms are non-vanishing, those corresponding respectively to the commutators $ [a(\bm{p},\lambda), a^\dagger(\bm{p}',\lambda')]$ and $ [a^\dagger(\bm{p},\lambda), a(\bm{p}',\lambda')]$ and one gets:
\begin{eqnarray}
&&\left[ \partial_{0}  {\Phi}(\bm{r},y,t) ,{\Phi}(\bm{r}^{\prime},y^{\prime},t)  \right]
    =  \frac{(-i)}{2}  \int \frac{d^3 \bm{p}}{(2\pi)^3}\, \int \frac{d^3 \bm{p}^{\prime}}{(2\pi)^3}\,
    \intop_0^{+\infty} d\lambda\, \mu(\lambda)  \intop_0^{+\infty} d\lambda^{\prime}\, \mu(\lambda^{\prime}) \times \nonumber \\
    &&
     \, (2\pi)^3\delta^3(\bm{p}-\bm{p}') \frac{1}{\mu(\lambda)}\delta(\lambda-\lambda')\, 
    \left[
    \psi(y,\lambda) \, \psi^*(y^{\prime},\lambda^{\prime})
    e^{i\bm{p}\cdot (\bm{r} - \bm{r}^{\prime})}\, e^{-i\omega_\lambda^2(\bm{p},\lambda)t}\, e^{+i\omega_{\lambda'^2}(\bm{p'}\!,\lambda')t}\right.\phantom{xxxxx} \\
    &&\left. \phantom{xxxxxxxxxxxxxxxxxxxxx}+
     \psi^*(y^{\prime},\lambda^{\prime})\, \psi(y,\lambda) \,
    e^{-i\bm{p}^{\prime}\cdot(\bm{r} - \bm{r}^{\prime})}\,e^{+i\omega_\lambda^2(\bm{p},\lambda)t}\, e^{-i\omega_{\lambda'^2}(\bm{p'}\!,\lambda')t} 
    \right].\nonumber
\end{eqnarray}
Finally, eliminating  the integrals  $\int d^3\bm{p}'$ and $ \int d\lambda'$ thanks to  the Dirac delta functions one is left with: 
\begin{eqnarray}
&&\left[ \partial_{0}  {\Phi}(\bm{r},y,t) ,{\Phi}(\bm{r}^{\prime},y^{\prime},t)  \right]
    =  \frac{(-i)}{2}  \int \frac{d^3 \bm{p}}{(2\pi)^3}\,\intop_0^{+\infty} d\lambda \, \mu(\lambda) 
    \left[
    \psi(y,\lambda) \, \psi^*(y^{\prime},\lambda)
    e^{i\bm{p}\cdot (\bm{r} - \bm{r}^{\prime})}\, \right.\phantom{xxxxx} \\
    &&\left. \phantom{xxxxxxxxxxxxxxxxxxxxxxxxxxxxxxxxxxxxxxxx}+
     \psi^*(y^{\prime},\lambda)\, \psi(y,\lambda) \,
    e^{-i\bm{p}^{\prime}\cdot(\bm{r} - \bm{r}^{\prime})} 
    \right].\nonumber
\end{eqnarray}
Upon use of the completeness relations (\ref{completeness}),   this becomes 
\begin{equation}
  \left[ \partial_{0}  {\Phi}(\bm{r},y,t) ,{\Phi}(\bm{r}^{\prime},y^{\prime},t)  \right]
    =  \frac{(-i)}{2} \delta^3(\bm{r}-\bm{r'}) \, \left[ y^{\alpha-1}\,\delta(y-y') + (y')^{\alpha-1}\,\delta(y'-y) \right] \, .
\end{equation}
This can be rewritten as:
\begin{equation}
\label{initial_condition_1}
  \left[ y^{1-\alpha}\,\partial_{0}  {\Phi}(\bm{r},y,t) ,{\Phi}(\bm{r}^{\prime},y^{\prime},t)  \right]
    =  -i \delta^3(\bm{r}-\bm{r'}) \,  \delta(y-y')
\end{equation}
which is, on account of Eq.~\eqref{conjugate_momentum_def},  the standard canonical equal time commutation relation between the field's conjugate momentum $\Pi  (\bm{r},y,t)$ and the field  $\Phi  (\bm{r},y,t)$ itself: 
\begin{equation}
\label{ETCRd+1}
  \left[ \Pi  (\bm{r},y,t) ,{\Phi}(\bm{r}^{\prime},y^{\prime},t)  \right]
    =  -i \delta^3(\bm{r}-\bm{r'}) \,  \delta(y-y')\,.
\end{equation}
This warrants that our $(4+1)$-dimensional theory with one extra space dimension is a standard local quantum field theory with canonical quantization. 
We emphasize how the possibility of carrying out the canonical quantization rests upon the completeness relation, Eq.~(\ref{completeness}) satisfied by the eigen-functions of the equation of motion along the extra-dimension with the appropriate function $\mu(\lambda)$.
Note that we cannot take the limit of Eqs.~\eqref{ETCRd+1},\eqref{initial_condition_1} on the brane as we know that in the nonlocal  $(3+1)$-dimensional theory the very concept of  conjugate momentum field is ill defined.

\subsection{Causal structure}
Let us consider now the  commutator, at different space-time points, of the bulk fields, $\left[ 
{\Phi}(x_{1},y_{1}), {\Phi}(x_{2},y_{2})
\right]$. This quantity can be computed using the field expansion in Eq.~\eqref{eq:def-Phi} and the quantization rules in Eq.~\eqref{quantization-rules1a}-\eqref{quantization-rules1c}.   
Then making use of the identity:
\begin{equation}
\label{relation}
\frac{1}{2 \omega_{\lambda^2}(\bm{p})} = \int dp_{0} \, \Theta(p_0) \, \delta \left( \bm{p}^2 - \lambda^2  \right)
\end{equation}
we find:
\begin{eqnarray}
\left[ 
{\Phi}(x_{1},y_{1}), {\Phi}(x_{2},y_{2})
\right] &=& \int d \lambda \, \mu (\lambda) \psi(y_{1},\lambda) \psi^*(y_{2},\lambda)
\int \frac{d^3 \bm{p}}{(2 \pi)^3}\, dp_0 \, \theta(p_0)\, \delta \left( \bm{p}^2 - \lambda^2 \right) \times \nonumber \\
&& 
\left[
 e^{ + i\left[ \bm{p} \cdot (\bm{x}_{1} - \bm{x}_{2}) 
 - p_0 (t_1 - t_2)\right] }
-e^{-i\left[ \bm{p} \cdot (\bm{x}_{1} - \bm{x}_{2}) 
 - p_0 (t_1 - t_2)\right] }
\right]
\end{eqnarray}
where $x_1=(t_1,\bm{x}_1)$ and $x_2=(t_2,\bm{x}_2)$. 
Recalling the definition of the Pauli-Jordan function:
\begin{equation}
    D_{\lambda^2}(x) = i \int \frac{d^4p}{(2\pi)^4} e^{-ip\cdot x} \, 2\pi \, \delta(p^2 -\lambda^2)\, \left[\theta(p^0) -\theta(-p^0) \right]
\end{equation}
we finally obtain,
\begin{equation}
    \left[ 
{\Phi}(x_{1},y_{1}), {\Phi}(x_{2},y_{2})
\right] = (-i) \intop_0^{+\infty}\, d\lambda\, \mu(\lambda)\, \psi^*(y_1, \lambda)\, \psi(y_2,\lambda) \, D_{\lambda^2}(x_1-x_2).
\label{field_commutator}
\end{equation}
Local commutativity  implies that the above field commutator for bulk space-like intervals between the points $(x_1,y_1)$ and 
$(x_2,y_2)$, would vanish identically. As also discussed in \cite{Mintchev:2001aa} for general functions $\mu(\lambda)$ this is not  the case in Eq.~\eqref{field_commutator}. Therefore in spite of the fact that the bulk field is built out of a classical \emph{local} action at the quantum level the field is nonlocal (local commutativity is lost). The origin of this very peculiar property is ultimately due the defect introduced in the bulk by the presence of the brane. The fact that the function $\mu(\lambda)$ appears in the quantization rules in Eq.~\eqref{quantization-rules1} shows that this is a genuine quantum phenomenon.

Taking the brane limit of Eq.~\eqref{field_commutator} on account of the boundary condition $\psi(y,\lambda) \to 1 $ as $y\to0$ we find:
\begin{equation}
    \left[\phi(x_1),\phi(x_2)\right] = -i \intop_0^{+\infty}\, d\lambda\, \mu(\lambda)\,  D_{\lambda^2}(x_1-x_2)\, .
\end{equation}
For space-like separations, $(x_1-x_2)$, the Pauli-Jordan function $D_{\lambda^2}(x_1-x_2)$ vanishes identically for any $\lambda$ (for it is the Pauli-Jordan function of a scalar field of mass $\lambda$) and therefore the field commutator will vanish. Thus the brane field satisfies local commutativity. On the other end the brane field is a non-canonical quantum field in the sense that it is associated to a nonlocal classical action and it cannot be quantized canonically. 
Indeed we have seen that the brane limit of the (canonical) equal-time commutation relation in the bulk, Eq.~\eqref{initial_condition_1}, is ill defined.

\subsection{Two-point correlation functions --the Wightman function and the Feynman propagator-- from the extension problem \label{subsec:unparticle}}

The nonlocal action in Eq.~\eqref{NLaction} as well as the e.o.m. in Eq.~\eqref{fraclaplaceeq} seems to be  
directly related to the unparticle propagator introduced in \cite{Georgi:2007ek} in the context of the study of a massive scale invariant theory. This can be easily seen by considering the Green function of Eq.~\eqref{fraclaplaceeq} defined in the usual way:
\begin{equation}
   (-\Delta)^\frac{\alpha}{2} G(x,x') =\delta^4(x-x')
    \label{fracGF}  
\end{equation}
The above equation is straightforwardly solved in Fourier space using \eqref{DeltaFourier},
\begin{equation}
     (-\Delta)^\frac{\alpha}{2} G(x-x') = \int \frac{d^4p}{(2\pi)^4}\, (p^2)^\frac{\alpha}{2}\, G(p)\,e^{ip\cdot (x-x')}\,.
\end{equation}
Then from Eq.~\eqref{fracGF} we obtain:
\begin{equation}
\label{fracprop}
    G(p) = (p^2)^{{-\alpha}/{2}}\, = \, \left({\sqrt{p^2}}\right)^{-\alpha} .
\end{equation}
This is exactly the unparticle propagator discussed for instance in Georgi's paper~\cite{Georgi:2007si} with $\alpha/2=2 -d_{\cal U} $ (up to an irrelevant numerical constant),
where $d_{\cal U}$ is the scaling dimension of the unparticle operator in the low-energy theory. Indeed, the unparticle propagator is obtained requiring 
scale invariance in the low-energy effective field theory of a theory with a nontrivial scale invariant IR fixed point \cite{Georgi:2007ek}.
Therefore, the unparticle propagator seems to be a feature of a nonlocal action of the type in Eq.~\eqref{NLaction}.

The particle content of the theory described by the fractional propagator in Eq.~(\ref{fracprop}) can be derived by computing a spectral density representation in terms of the discontinuities across the branch cut. For any analytic function $f(z)= \int_{-\infty}^{+\infty}dt\, \rho(t) \frac{1}{z-t}$ the discontinuity across the branch cut is: $\text{disc}[f(z)]=\left\{f(z+i\epsilon) -f(z-i\epsilon)\right\}_{\epsilon\to 0} = - 2\pi i \rho(z)$. 
We define the argument of the complex $z$-variable, $\theta$,  to vary in the interval $-\pi < \theta < \pi$ so that the funztion $\sqrt{z}$ has a branch cut along the negative (real) $z$-axis. We therefore let $z=-p^2$ and the propagator in Eq.~(\ref{fracprop}) and the fractional propagator $(\sqrt{-z})^{-\alpha}$ will have a discontinuity for positive $z$. Explicitly we find:
\begin{equation}
\text{disc}\left[(\sqrt{-z})^{-\alpha}\right] = \left\{\begin{array}{ll}
2i \sin(\frac{\alpha\pi}{2})\left(\sqrt{z}\right)^{-\alpha}&z>0\\
0&z\le0
\end{array} \right. 
\end{equation}
We therefore can extract the density $\rho(z)$ and finally obtain the spectral representation: 
\begin{equation}
\frac{1}{\left(\sqrt{-z}\right)^\alpha}= \int_0^{+\infty} \!\! d t\, \frac{1}{\pi} \sin\left(\frac{\pi \alpha}{2}\right) \left(\sqrt{t}\right)^{-\alpha}  \, \frac{1}{-z + t} \, . 
\end{equation}
Going back to the $p^2$ variable the fractional propagator can then be written as $t \to m^2$:
\begin{equation}
\label{spectral_frac_propagator}
\frac{1}{\left(p^2\right)^{\alpha/2}}= \int_0^{+\infty} \!\! d m\, \frac{2}{\pi} \sin\left(\frac{\pi \alpha}{2}\right) m^{1-\alpha}  \, \frac{1}{p^2 + m^2} \, . \end{equation}
and we obtain that the fractional propagator is a continuous distribution of scalar massive propagators with mass parameter $m$ weighed by a spectral density $\propto m^{1-\alpha}$.   It is easily shown that in the limit $\alpha \to 2^-$ (or $d_{\cal U}\to 1^+$) the right hand side of Eq.~\eqref{spectral_frac_propagator} reduces to the massless scalar propagator $1/p^2$ as upon putting $\alpha =2-\epsilon$ one obtains  $\left[\frac{2}{\pi} \sin\left(\frac{\pi \alpha}{2}\right) m^{1-\alpha}\right]_{\alpha=2-\epsilon} \to \epsilon\, m^{-1 +\epsilon} \to 2\, \delta(m)$ on account on account of the Dirac $\delta$-function representation $\delta(x)= \lim_{\epsilon\to0} \frac{\epsilon}{2}\, |x|^{-1+\epsilon}$ (the factor 2 is canceled by the fact that the integral in $dm$ ranges only over half the real axis).

We will show explicitly in the following that it is possible to obtain the same result for the propagator starting from the $(d+1)$ local theory.

Using the field decomposition in~\eqref{eq:def-Phi} we now compute the bulk Wightman function for the field ${\Phi}(x,y)$  
\begin{eqnarray}
   \langle 0| {\Phi}(x,y)&&\!\!\!{\Phi}(x',y')|0 \rangle  = \frac{1}{2}\int \frac{d^3 \bm{p}}{(2\pi)^3}\, \int d\lambda\, \mu(\lambda) \int \frac{d^3 \bm{p'}}{(2\pi)^3}\,\int  d\lambda'\, \mu(\lambda') 
   \times\\  
   && \phantom{\!\!\!\!\!}\langle 0 | \left[ a(\bm{p},\lambda) \frac{e^{i[\bm{p}\cdot\bm{x}-\omega_{\lambda^2}(\bm{p})t]}}{\sqrt{\omega_{\lambda^2}(\bm{p})}} \psi(y,\lambda)  +a^\dagger(\bm{p},\lambda) \frac{e^{-i[\bm{p}\cdot\bm{x} -\omega_{\lambda^2}(\bm{p})t]}}{\sqrt{\omega_{\lambda^2}(\bm{p})}} \psi^*(y,\lambda)\right] \times\nonumber\\
   && \phantom{\!\!\!\!\!}\left[ a(\bm{p}',\lambda') \frac{e^{i[\bm{p}'\cdot\bm{x}'-\omega_{\lambda'^2}(\bm{p'})t']}}{\sqrt{\omega_{\lambda'^2}(\bm{p'})}} \psi(y',\lambda')  +a^\dagger(\bm{p'},\lambda') \frac{e^{-i[\bm{p'}\cdot\bm{x'} -\omega_{\lambda'^2}(\bm{p'})t']}}{\sqrt{\omega_{\lambda'^2}(\bm{p'})}} \psi^*(y',\lambda')\right] |0\rangle. \nonumber
\end{eqnarray}
Expanding the product of the two factors in square parenthesis, when taking the vacuum expectation value, we get a nonzero contribution only from the cross term of the type $a(\bm{p},\lambda) a^\dagger(\bm{p}',\lambda')$ :
\begin{eqnarray}
   \langle 0| {\Phi}(x,y)&&\!\!\!{\Phi}(x',y')|0 \rangle  = \frac{1}{2}\int \frac{d^3 \bm{p}}{(2\pi)^3}\, \int \frac{d^3 \bm{p'}}{(2\pi)^3}\, \int d\lambda\, \mu(\lambda)\int  d\lambda'\, \mu(\lambda') 
   \times\\  
   && \langle 0 |  a(\bm{p},\lambda) \, a^\dagger(\bm{p}',\lambda') |0\rangle \,  \frac{e^{+i[\bm{p}\cdot\bm{x} -\omega_{\lambda^2}(\bm{p})t]}}{\sqrt{\omega_{\lambda^2}(\bm{p})}}\, \frac{e^{-i[\bm{p}'\cdot\bm{x}'-\omega_{\lambda'^2}(\bm{p'})t']}}{\sqrt{\omega_{\lambda'^2}(\bm{p'})}} \psi^*(y,\lambda) \psi(y',\lambda') \nonumber
\end{eqnarray}
and the vacuum expectation value  $\langle 0| a(\bm{p},\lambda) \, a^\dagger(\bm{p}',\lambda')  |0\rangle $ can be computed from the identity:
\begin{eqnarray}
 \langle 0| a(\bm{p},\lambda) \, a^\dagger(\bm{p}',\lambda')  |0\rangle  &= &\langle 0|\left[ a(\bm{p},\lambda) , a^\dagger(\bm{p}',\lambda')\right] +    a^\dagger(\bm{p}',\lambda') \, a(\bm{p},\lambda) |0\rangle   \nonumber \\
 &=& \langle 0|\left[ a(\bm{p},\lambda) , a^\dagger(\bm{p}',\lambda')\right]  |0\rangle .
\end{eqnarray}
Using the quantization rules given in Eqs.~\eqref{quantization-rules1a}-\eqref{quantization-rules1c} we find:
\begin{equation}
\label{aadaggervev}
 \langle 0| a(\bm{p},\lambda) \, a^\dagger(\bm{p}',\lambda')  |0\rangle =  (2\pi)^3\delta^3(\bm{p}-\bm{p}') \frac{1}{{\mu}(\lambda)} \delta(\lambda -\lambda')  
\end{equation}
and, using the result in Eq.~\eqref{aadaggervev}, we find
\begin{eqnarray}
   \langle 0| {\Phi}(x,y)&&\!\!\!{\Phi}(x',y')|0 \rangle  = \frac{1}{2}\int \frac{d^3 \bm{p}}{(2\pi)^3}\, \int \frac{d^3 \bm{p'}}{(2\pi)^3}\, \int d\lambda\, \mu(\lambda)\int  d\lambda'\, \mu(\lambda') 
   \times\\  
   &&  \frac{e^{+i[\bm{p}\cdot\bm{x} -\omega_{\lambda^2}(\bm{p})t]}}{\sqrt{\omega_{\lambda^2}(\bm{p})}}\, \frac{e^{-i[\bm{p}'\cdot\bm{x}'-\omega_{\lambda'^2}(\bm{p'})t']}}{\sqrt{\omega_{\lambda'^2}(\bm{p'})}} \psi^*(y,\lambda) \psi(y',\lambda')\nonumber \times \\ && (2\pi)^3\delta^3(\bm{p}-\bm{p}') \frac{1}{{\mu}(\lambda)} \delta(\lambda -\lambda')\, .      \nonumber
\end{eqnarray}
Then, performing the integrations in $\bm{p}'$ and $\lambda'$ by making use of the delta functions we obtain:
\begin{eqnarray}
 \langle 0| {\Phi}(x,y)&&\!\!\!{\Phi}(x',y')|0 \rangle  =    \int\!\!\! \frac{d^3 \bm{p}}{(2\pi)^3}  \!\!\!\intop_0^{+\infty}\!\!\! d\lambda \, \mu(\lambda)
 \, \frac{e^{i\bm{p}\cdot(\bm{x}-\bm{x}')} 
 e^{-i\omega_{\lambda^2}(\bm{p}) (t-t')}}{2 \,\omega_{\lambda^2}(\bm{p})}
 \, \psi^*(y,\lambda) \psi(y',\lambda). \qquad
\end{eqnarray}
The integral over $\bm{p}$ gives, on account of Eq.~\eqref{relation}, the Wightman function~\cite{Peskin:1995ev} of a scalar field of mass $\lambda$,
\begin{eqnarray}
\label{Wightman_massive}
    W_{\lambda^2} (x,x') &=& \int\, \frac{d^3\bm{p}}{(2\pi)^3} \, \frac{e^{-i p\cdot (x-x')}}{2\, \omega_{\lambda^2}(\bm{p})}\nonumber \\
    &=& \int \frac{d^4p}{(2\pi)^4} \,\Theta (p_0)\, 2\pi\, \delta(p^2-\lambda^2)\, e^{-i p\cdot (x-x')}
\end{eqnarray}
where $W_{\lambda^2}(p)= \Theta(p^0)\, 2\pi\,\delta(p^2-\lambda^2) $ is the Fourier-space Wightman function of mass $\lambda$.
The bulk two-point function can therefore be written as
\begin{equation}
 \langle 0| {\Phi}(x,y){\Phi}(x',y')|0 \rangle  =     \int_0^\infty d\lambda \, \mu(\lambda)\, \, W_{\lambda^2}(x,x')  \, \psi^*(y,\lambda) \psi(y',\lambda)\,.
\end{equation}
The brane induced Wightman function of the nonlocal field $\phi(x)$, on account of the boundary conditions imposed on the wave-functions, $\psi(y,\lambda)|_{y=0}=1 $, is
\begin{equation}
\label{two-point-Wightman}
 \langle 0| {\phi}(x){\phi}(x')|0 \rangle  =  \lim_{y,y'\downarrow 0}   \langle 0| {\Phi}(x,y){\Phi}(x',y')|0 \rangle  = \int_0^\infty \!\!\!d\lambda \, \mu(\lambda)\,  \, W_{\lambda^2}(x,x')
\end{equation}
Therefore we can see that the two-point Wightman function for the brane field can be expressed as an integral over the mass of a Wightman function of an ordinary  massive scalar field.

We now make contact  between our approach and that of refs.~\cite{Georgi:2007si,Georgi:2007ek}
which fixes the normalization of a scalar field $O_{\cal U}(x)$ of scaling dimension $d_{\cal U}$ by comparison with the phase space of $n$ massless-particles. Indeed the correlation function for the field   $O_{\cal U}(x)$ can be written in terms of a spectral density $\rho(P^2)$:
\begin{equation}
\label{Wightman_Georgi_1}
\langle 0| O_{\cal U}(x) O^\dagger_{\cal U}(0) |0 \rangle = \int \, \frac{d^4 P}{(2\pi)^4}\, e^{-iP\cdot x} |\langle0| O_{\cal U}(0)|p\rangle|^2 \rho(P^2)
\end{equation}
and on dimensional grounds, given that $O_{\cal U}$ has scaling dimension $d_{\cal U}$ one can write:
\begin{equation}
\label{Wightman_Georgi_2}
 |\langle0| O_{\cal U}(0)|p\rangle|^2 \rho(P^2) = A_{d_{\cal U}}\,  \theta(P^0)\,\theta(P^2) \,(P^2)^{d_{\cal U}-2} 
\end{equation}
with \begin{equation}
 A _ { d_{\cal U} } = \frac { 16 \pi ^ { 5 / 2 } } { ( 2 \pi ) ^ { 2 d_{\cal U} } } \frac { \Gamma ( d_{\cal U} + 1 / 2 ) } { \Gamma ( d_{\cal U} - 1 ) \Gamma ( 2 d_{\cal U} ) }  %
\end{equation}
a numerical constant that has been related in \cite{Georgi:2007si,Georgi:2007ek} to the phase space factor of a fractional number ($d_{\cal U}$) of massless particles.

The field $O_{\cal U}(x)$ must coincide, up to a numerical constant, with the field  defined within our approach, \`{a} la Caffarelli-Silvestre, by  $\phi(x) =\lim_{y\to 0^+}\Phi(x,y)$, with $\alpha= 4-2d_{\cal U}$ so that we can put:
\begin{equation}
    \phi(x) = C'\, O_{\cal U}(x)\, .
    \label{field_normalization}
\end{equation}
By using Eq.~\eqref{Wightman_massive} into  Eq.~\eqref{two-point-Wightman} we can write the two-point Wightman function of the (neutral) field $\phi(x)$ as:
\begin{equation}
    \langle 0| \phi(x) \phi(0) |0 \rangle = \int \frac{d^4P}{(2\pi)^4} \, e^{-iP\cdot x} \, 2\pi \,\theta(P^0)\, \theta(P^2)\, \frac{\mu(\sqrt{P^2})}{2\sqrt{P^2}}\, .
\end{equation}  Then making use of the explicit expression of the function $\mu(\lambda)$ given in Eq.~\eqref{mufuntion} and comparing with Eqs.~\eqref{Wightman_Georgi_1}\&\eqref{Wightman_Georgi_2} allows us to extract the relation between the constant $C'$ and the quantity $A_{d_{\cal U}}$: 
\begin{equation}\left.\left[ \frac{2^{-\alpha/2}\pi}{\Gamma(\alpha/2) \sin(\pi\alpha/2)} \right]^{-2}\right|_{\alpha=4 -2d_{\cal U}} = (C')^2\,\frac{1}{\pi} \, A_{d_{\cal U}}
\label{Constant_Georgi}
\end{equation}
Numerically it is easily found that:
\begin{equation}
\label{constantCprime}
    (C')^2 = {(2\pi)^{2-\alpha}}\, \left(2-\alpha\right)
\end{equation}

We note that the same reasoning leading to Eq.~\eqref{two-point-Wightman} would allow to establish the same identity for the time ordered propagator.

Indeed, the time-ordered product of the local fields is
\begin{eqnarray}
    \langle 0| T \big[ {\Phi}(x,y){\Phi}(x',y') \big] |0 \rangle  &=&     \int_0^\infty d\lambda \, \mu(\lambda) \, \psi^*(y,\lambda) \psi(y',\lambda)\, \times  \\
    && \int\!\! \frac{d^{3} \bm{p}}{(2 \pi)^3} \frac{e^{-i p \cdot \left( x - x^{\prime}\right)}}{2 \omega_{\lambda^2} (\bm{p})} \left[ e^{- i \omega_{\lambda^2} (\bm{p}) \tau} \theta \left( t- t^{\prime} \right) + e^{+ i \omega_{\lambda^2} (\bm{p}) \tau} \theta \left( t^{\prime} - t ) \right) \right]\nonumber 
\end{eqnarray}
and using the integral representation of the Heaviside $\theta$ function \begin{eqnarray}
    \langle 0| T \big[ {\Phi}(x,y){\Phi}(x',y') \big] |0 \rangle  &=&     \int_0^\infty d\lambda \, \mu(\lambda) \, \psi^*(y,\lambda) \psi(y',\lambda)\, \times \nonumber \\
    &&\phantom{xxxxxxxx} \lim_{\epsilon \rightarrow 0} \int \frac{d^3 \bm{p}}{(2 \pi)^3} \frac{d \omega}{2 \pi } \frac{i\, e^{i \big[ \omega (t- t^{\prime}) - p \cdot (x - x^{\prime})\big] }}{\omega^2 - \omega^{2}_{\lambda^{2}} +i \epsilon } 
\end{eqnarray} 
or, using the momentum four-vector $k=(\omega, \bm{p})$
\begin{equation}
    \langle 0| T \big[ {\Phi}(x,y){\Phi}(x',y') \big] |0 \rangle  = \lim_{\epsilon \rightarrow 0}    \int_0^\infty d\lambda \, \mu(\lambda) \, \psi^*(y,\lambda) \psi(y',\lambda)\,    \int \frac{d^4 k}{(2 \pi)^4}  \frac{i \, e^{i k \cdot  (x- x^{\prime})  }}{k^2 - \lambda^{2} +i \epsilon }.\qquad 
\end{equation} 
Therefore, we get
\begin{equation} \label{eq:propLoc}
    \langle 0| T \big[ {\Phi}(x,y){\Phi}(x',y') \big] |0 \rangle  = 
    \int_0^\infty d\lambda \, \mu(\lambda) \, \psi^*(y,\lambda) \psi(y',\lambda)\, D_{F} \left( x - x^{\prime}, \lambda^{2} \right)
\end{equation}
where we have introduced the Feynman propagator of a scalar field of mass $m$
\begin{equation}
    \label{eq:propLoc2}
    D_{F} \left( x - x^{\prime}, \lambda^{2} \right) = \lim_{\epsilon \rightarrow 0} \int \frac{d^4 k}{(2 \pi)^{4}} \frac{i\, e^{i k \cdot (x - x^{\prime})}}{k^2 - \lambda^2 + i \epsilon}.
\end{equation}
Now, from Eq. \eqref{eq:propLoc} we can extract the propagator for the nonlocal theory by taking the brane limit that defines the fields $\phi (x) = \lim_{y \rightarrow 0} \Phi (x, y)$, therefore
\begin{eqnarray}
    \langle 0| T \big[ {\phi}(x){\phi}(x') \big] |0 \rangle  = \lim_{y,y^{\prime} \rightarrow 0} \langle 0| T \big[ {\Phi}(x,y){\Phi}(x',y') \big] |0 \rangle  
\end{eqnarray}
Since we have that the wave-functions in \eqref{eq:propLoc2} satisfy the boundary conditions \eqref{eq:bcy}, we finally have
\begin{equation}
\label{final_prop}
    \langle 0| T \big[ {\phi}(x){\phi}(x') \big] |0 \rangle  = \int^{\infty}_{0} d \lambda \, \mu (\lambda)  D_{F} \left( x - x^{\prime}; \lambda^{2} \right)
\end{equation}
Now, using  \eqref{mufuntion}
and  taking the Fourier transform of Eq.~\eqref{final_prop}, we can find the momentum space propagator $D(p)$  of the brane nonlocal field, $\phi(x)$, as:
\begin{equation}
\label{prop_nonlocal}
    D(p) = \frac{i}{2} \left[  \frac{2^{-\frac{\alpha}{2}} \pi}{ \Gamma (\frac{\alpha}{2}) \sin \left(\pi \frac{\alpha}{2} \right)} \right]^{-2} \int_0^\infty \, d\lambda^2 \, (\lambda^2)^{-{\alpha}/{2}}\, \frac{1}{p^2 -\lambda^2 +i\epsilon}\, .
\end{equation}
We remark first of all that if Wick rotated the  result of Eq.~(\ref{prop_nonlocal}) coincides,  up to an immaterial normalization constant, with the result  obtained in subsection \ref{subsec:unparticle} (see Eq.~\eqref{spectral_frac_propagator}) for the fractional propagator: i.e. the two propagators are characterized by the same particle content.
Eq.~(\ref{prop_nonlocal}) is then to be compared with the propagator of the unparticle field of dimension $d_{\cal U}$~\cite{Georgi:2007si} given  as  \footnote{See eq. 3 in~\cite{Georgi:2007si}.}:
\begin{equation}
\label{prop_Georgi}
D_U(p) = i \,\frac{A_{d_{\cal U}}}{2 \pi} \int_0^\infty dM^2 \, (M^2)^{d_{\cal U}-2}\, \frac{1}{p^2 -M^2 + i\epsilon}
\end{equation}
By comparing Eq.~\eqref{prop_nonlocal} to Eq.~\eqref{prop_Georgi} in order for the two propagators   to satisfy $D(p) = (C')^2 D_U(p) $, which follows from Eq.~(\ref{field_normalization}), we must have, for the constant $(C')^2$, again Eq.~\eqref{Constant_Georgi}.
The constant $(C')^2$ can be computed on account of the relation $\alpha=4-2d_{\cal U}$ and we find again, as expected,  the same result of Eq.~\eqref{constantCprime}.
Computing explicitly  the integrations in Eq.~\eqref{prop_nonlocal} and Eq.~\eqref{prop_Georgi} we find:
\begin{subequations}
\label{explcit_propagator}
 \begin{align}
 \label{explcit_propagator_a}
  D(p) & = - i\,
   \frac{2^{\alpha -1} \sin \left(\frac{\pi  \alpha }{2}\right) 
   \Gamma \left(\frac{\alpha }{2}\right)^2}{\pi ^2}\,
    \left( -p^2 -i \epsilon \right)^{-\frac{\alpha}{2}}\, , \\
    \label{explcit_propagator_b}
 D_{\cal U}(p)  & = + i \,\frac{A_{d_{\cal U}}}{2 \sin(\pi d_{\cal U})} \,(-p^2-i\epsilon)^{d_{\cal U}-2}\, ,
 \end{align}
\end{subequations}
which indeed differ exactly by the constant $(C')^2$ as in Eq.~\eqref{constantCprime}. Note that while  Eq.~\eqref{explcit_propagator_b} in the limit $d_{\cal U} \to 1$ reproduces exactly the mass-less scalar field propagator  $D_{\cal U}(p) \to {+i}/{(p^2 +i\epsilon)}$, Eq.~\eqref{explcit_propagator_a} when  $\alpha\to 2$ reproduces the same mass-less scalar propagator only up to the numerical constant $(C')^2$: $D(p)\to {+i (2-\alpha)}/{(p^2+i\epsilon)}$. It is thus clear the advantage of working with the normalization introduced by Georgi~\cite{Georgi:2007ek, Georgi:2007si} for  the nonlocal unparticle field.   

We conclude this section by showing explicitly how the extension problem straightforwardly provides the  standard unparticle effective action routinely used in the phenomenology of this model.
We start then from the central result of the extension problem, namely Eq.~\eqref{eq:nonlocTOloc2}, which relates the local action in (4+1) dimension to a nonlocal action in 4 dimensions:
\begin{equation}
\label{final_non_local_action}
    S= \frac{1}{2C} \, \int d^4 x\, \phi(x)\, \Box^{\alpha/2}\, \phi(x)\, .
\end{equation}
Here we redefine the field 
using Eq.~\eqref{field_normalization} and use the field $O_{\cal U}(x)$ normalized ``\`{a} la Georgi'', i.e. to have the same propagator computed in \cite{Georgi:2007si}. We then obtain:
\begin{equation}
    S= \frac{(C')^2}{C} \,\frac{1}{2} \int d^4 x\, O_{\cal U}(x)\, \Box^{\alpha/2}\, O_{\cal U}(x)\, .  
\end{equation}
Then by using the explicit results derived above, c.f. Eq.~\eqref{Constant_Georgi}, together with Eq.~\eqref{eq:preConst} and upon the replacement $\alpha= 4-2d_{\cal U}$ we find:
\begin{equation}\label{unparticle_effective_action}
    S= \frac{2\sin(\pi d_{\cal U})}{A_{d_{\cal U}}} \,\frac{1}{2} \int d^4 x\, O_{\cal U}(x)\, \Box^{2-d_{\cal U}}\, O_{\cal U}(x)\, .  
\end{equation}
This classical action being quadratic in the field is equivalent to the effective action and thus Eq.~\eqref{unparticle_effective_action} is the effective action for unparticles. Indeed its second order funtional derivative with respect to the fields gives the inverse two point function according to:
\begin{equation}
    \frac{\delta S}{\delta O_{\cal U}(x) \delta O_{\cal U}(x')} = i D^{-1}_{\cal U}(x,x')= \frac{2\sin(\pi d_{\cal U})}{A_{d_{\cal U}}}\,\Box^{2-d_{\cal U}}
\end{equation}
which is consistent with the explicit computation of the propagator offered above c.f. Eq.~\eqref{explcit_propagator_b}.  
\section{Vacuum energy density}

Here we discuss the brane vacuum energy of the nonlocal theory  as obtained from the local Hamiltonian of the extended theory in  ($d$+1) dimensions. The bulk Hamiltonian operator ${\cal H}$ can be obtained as the component $T_{00}$ of the energy momentum tensor $T_{\mu\nu}$ of the $(d+1)$ local theory.
\begin{equation}
   {\cal H}= \frac{1}{2} \int d^3\bm{x} \intop_0^{+\infty} dy\, y^{1-\alpha}\left[\partial_0\Phi(x,y)\, \partial_0\Phi(x,y) -\Phi(x,y)\,\partial_0^2\Phi(x,y) \right]
\end{equation}
with $x=(\bm{x},t)$.
We start again from the field expansion in Eq.~\eqref{eq:def-Phi} and calculate the bulk energy as ${ E}^{\text{bulk}}_{\text{vac}} = \langle 0| {\cal H} |0 \rangle$ that gives:
\begin{eqnarray}
 { E}^{\text{bulk}}_{\text{vac}} 
 &=& \frac{1}{2} \int d^3\bm{x}\, \int \frac{d^3\bm{p}}{(2\pi)^3} \intop_0^{+\infty}\, d\lambda\, \mu(\lambda)\, \omega_{\lambda^2}(\bm{p}) \intop_0^{+\infty} dy\, y^{1-\alpha} \psi^*(y,\lambda)\, \psi(y,\lambda)
 \label{eq:bulk_energy}
\end{eqnarray}
It can easily be verified, by taking into account the dimensions of the function $\mu(\lambda)$ that the above  quantity is correctly dimensioned as an energy. From this bulk energy we ought to extract the vacuum brane energy ${E}_{\text{vac}}^{\text{brane}}$. One possibility is to think to slice up the extra space dimension $y \in [0,+\infty]$ and assign the first slice $y\in[0, y_\text{cut}]$ to the brane. Note that in Eq.~\eqref{eq:bulk_energy}  the integration over the 3-space entails an infinite factor. To avoid such infinite factor as it is standard in quantum field theory, we imagine enclosing our brane on a large but  finite volume $V$. This allows to define a brane vacuum  energy density:
\begin{equation}
    {\varepsilon}_\text{vac}=\frac{1}{2}  \int \frac{d^3\bm{p}}{(2\pi)^3} \intop_0^{+\infty}\, d\lambda\, \mu(\lambda)\, \omega_{\lambda^2}(\bm{p}) \intop_0^{y_\text{cut}} dy\, y^{1-\alpha} \psi^*(y,\lambda)\, \psi(y,\lambda)
 \label{eq:brane_energy_density}
\end{equation}
The cutoff length $y_\text{cut}$ corresponds  (in natural units) to an energy scale \begin{equation}
2\pi\Lambda_{\cal U} = 1/y_\text{cut}\, .    
\end{equation} This physical cutoff defines, on the brane, an effective field theory which corresponds to integrating out,  in Eq.~\eqref{eq:bulk_energy}, all the degrees of freedom at energies $E$ higher than $2\pi\Lambda_{\cal U}$ ($E\ge2\pi\Lambda_{\cal U}$). Indeed the $y$-integral in Eq.~\eqref{eq:brane_energy_density} can be  computed approximating the wave functions  by means of the imposed boundary conditions $\psi(y,\lambda) \to 1$ as $y \to 0$ and  converting it into  an energy integral (over the variable $E=1/y$):
\begin{equation}
    {\varepsilon}_\text{vac}= \intop_0^{+\infty}\, d\lambda\, \mu(\lambda)\, \int \frac{d^3\bm{p}}{(2\pi)^3}\,  \frac{\omega_{\lambda^2}(\bm{p})}{2} \intop_{2\pi \Lambda_{\cal U}}^{\infty} dE\, E^{\alpha-3} 
 \label{eq:brane_energy_density_EFT1}
\end{equation}
which shows clearly how the degrees of freedom $E\ge2\pi \Lambda_{\cal U}$ are being integrated out. This defines an effective field theory valid up to energies $\approx 2\pi \Lambda_{\cal U}$.  
The vacuum energy density on the brane is finally given by:
\begin{equation}
    {\varepsilon}_\text{vac}=\frac{1}{(2-\alpha)(2\pi \Lambda_{\cal U})^{2-\alpha}} \intop_0^{+\infty}\, d\lambda\, \mu(\lambda)\, \int \frac{d^3\bm{p}}{(2\pi)^3}\,  \frac{\omega_{\lambda^2}(\bm{p})}{2} 
 \label{eq:brane_energy_density_EFT2}\,.
\end{equation}

Note that if $d_{\cal U}$ is the scaling dimension of the field then $\alpha=4-2d_{\cal U}$ and $2-\alpha= 2 d_{\cal U} -2 >0$ for $d_{\cal U}>1$. We note that in Eq.~\eqref{eq:brane_energy_density_EFT2} the fractional power of $\Lambda_{\cal U}$ exactly compensates the fractional dimensions of the function $\mu(\lambda)\propto   \lambda^{1-\alpha}$ thus leaving a genuine energy density. As a final remark we note that Eq.~\eqref{eq:brane_energy_density_EFT2} expresses the  vacuum energy density of the nonlocal effective theory as an integral over the mass parameter $\lambda$ of the vacuum energy density of a scalar field of mass $\lambda$, since ${\omega_{\lambda^2}(\bm{p})}/{2} $ is the zero point energy of a mode of momentum $\bm{p}$ and mass $\lambda$, weighted by the function $\mu(\lambda)$. Eq.~\eqref{eq:brane_energy_density_EFT2} coincides up to a numerical constant with the central result of ref.~\cite{FRASSINO2017675} for the vacuum energy of unparticles:   it is still a divergent quantity but it can for instance be used to compute the unparticle Casimir energy by imposing appropriate geometrical boundary conditions~\cite{FRASSINO2017675}. 
Note that Eq.~\eqref{eq:brane_energy_density_EFT2} can be cast in the standard normalization routinely used for unparticles (c.f. \cite{Georgi:2007si,Georgi:2007ek}) by using the result in Eq.~\eqref{Constant_Georgi}. Namely:
\begin{equation}
    \mu(\lambda)
    = \,(C')^2\, \frac{A_{d_{\cal U}}}{\pi}\, \lambda^{1-\alpha} = (2\pi)^{2-\alpha} (2-\alpha)\,\frac{A_{d_{\cal U}}}{\pi}\, \lambda^{1-\alpha}
\end{equation}
and then by making use of Eq.~\eqref{constantCprime} we find ($2-\alpha\,=\, 2 d_{\cal U} -2 $):
\begin{equation}
    {\varepsilon}_\text{vac}=\frac{A_{d_{\cal U}}}{ \pi\, \Lambda_{\cal U}^{2d_{\cal U}-2}} \intop_0^{+\infty}\, d\lambda\, \lambda^{2d_{\cal U}-3}\, \int \frac{d^3\bm{p}}{(2\pi)^3}\,  \frac{\omega_{\lambda^2}(\bm{p})}{2} 
 \label{eq:brane_energy_density_EFT3}\,,
\end{equation} 
and this coincides 
with the central result derived in \cite{FRASSINO2017675} for the \emph{UnCasimir} effect.

\section{Negative powers of the fractional Laplacian \label{sec:negative-power}} 
In \cite{doi:10.1080/03605301003735680}, it has been shown that the extension problem in the original formulation of Eq. \eqref{eq:Delta} can be directly applied to the inverse powers of a generic second order differential operator (i.e., $\Delta^{-\alpha/2}$ with $\alpha>0$).
Interestingly, it has also been discussed in~\cite{2016arXiv160307988D} that the same extension problem  is valid for negative powers of the fractional Laplacian but mapping the Dirichlet boundary condition to a Neumann boundary condition. 
\\
In particular, for $\alpha \in (0,2)$ and $f: \mathbb{R}^n \rightarrow \mathbb{R}$, a smooth and bounded function, one can consider the following extension problem to the upper half space with
a Neumann-type boundary condition:
 \begin{subnumcases}{}
    \!\! \frac{\partial w}{\partial y^{\alpha}} (x,0)= - f(x) \qquad x \in \mathbb{R}^n, \label{eq:bc2} \\
    \!\! \nabla \cdot \left( y^{1-\alpha} \nabla w \right) = 0 \qquad x \in \mathbb{R}^n, y>0. \label{eq:eom2}
\end{subnumcases}
Note that in order to have a well-posed problem an extra
condition should be imposed to avoid the fact that if $w$ is a solution, then $w + c$ is also a
solution for any $c \in \mathbb{R} $. To this end, the following decay condition for large $y$ can be assumed:
\begin{equation}
    \lim_{y\rightarrow \infty} w(x,y) \rightarrow 0. \label{eq:bcc2}
\end{equation}
Then the solution $w(x,y)$ of problem \eqref{eq:bc2}-\eqref{eq:eom2} that satisfies the b.c. \eqref{eq:bcc2} has an explicit expression given by
\begin{eqnarray}
   S_{n,\alpha}\;(-\Delta)^{-\alpha/2}\;f(x)=\lim_{y\rightarrow 0^{+}} w(x,y) \label{eq:DeltaInv}
\end{eqnarray}
where
$S_{n,\alpha}$ is a constant defined as
\begin{eqnarray}
    S_{n,\alpha} = \frac{C_{n,2-\alpha}}{(n-\alpha)\; D_{n,\alpha}}
\end{eqnarray}
with
\begin{eqnarray}
    C_{n,2-\alpha}:= \frac{2^\alpha \; \Gamma \left( \frac{n+\alpha}{2} \right)}{\pi^{n/2} \Gamma \left( -\frac{\alpha}{2} \right)},\qquad 
    D_{n,\sigma}:= \frac{\pi^{\alpha - n/2}}{\Gamma \left( \frac{s}{2} \right)} \Gamma \left( \frac{n-s}{2} \right). 
\end{eqnarray}
Therefore, while the fractional Laplacian
is an operator that maps Dirichlet boundary conditions to Neumann-type conditions
for the ``local'' extended problem \eqref{eq:bc}-\eqref{eq:eom}; the inverse fractional Laplacian  is an
operator that maps a Neumann-type boundary condition to a Dirichlet condition
for the extended problem \eqref{eq:bc2}-\eqref{eq:eom2}, at least for $\alpha \in (0,2)$.

We show here how this can be explicitly done in terms of the eigenfunctions of the problem given in Eq.~\eqref{eq:psiBessel}.
The boundary conditions on the eigenfunctions $\psi(\lambda,y)$ now become:
\begin{equation}
   \psi(\lambda, 0) =0,\qquad y^{1-\alpha} \partial_{y} \left. \psi(\lambda, y) \right|_{y=0} =1. 
\end{equation}
which  from Eq.~\eqref{eq:psiBessellty0} can easily be shown to be solved by:
\begin{equation}
   A= \frac{2^\frac{\alpha}{2}\Gamma(\frac{\alpha}{2}+1)}{\alpha \lambda^\alpha}\,, \qquad \qquad  B= 0\,, 
\end{equation}
which lead, through Eq.~\eqref{eq:psiBessel}, to the solutions:
\begin{equation}
   \psi(\lambda , y) = 
   \frac{2^\frac{\alpha}{2}\Gamma(\frac{\alpha}{2}+1)}{\alpha \lambda^\alpha}\,
   (\lambda y)^{\frac{\alpha}{2}} J_{\alpha/2} (\lambda y). \label{eq:psiBessel4}
\end{equation}
It is important to note that, we can apply here the same procedure as before because the local bulk field theory satisfies the same equation of motion Eq. \eqref{eq:eom2}.

\section{Discussion and applications to GR \label{sec:discussion}}
In the previous sections we have discussed particular aspects of a scalar nonlocal quantum field theory. Some properties of the theory can be obtained thinking about the nonlocal fields as living on a brane and studying a local bulk theory with an extra-dimension. The features previously discussed can be applied to some extent to other fields and in particular,
a possible natural application would be to the so-called DGP (Dvali-Gabadadze-Porrati) model \cite{Dvali:2000hr}.
The DGP model is a possible way of describing the realization of a continuum of massive gravitons using extra-dimensions and a brane on which the Standard Model is confined.  \\
    In particular the DGP model describes a $(3+1)$-dimensional brane living in a $(4+1)$-dimensional bulk spacetime, and the total action has both a $5$d and a $4$d parts, that reads schematically
    \begin{equation}
        S= \frac{M^{3}_{5}}{2} \int d^5 X \, \sqrt{-G} \,R(G) + \frac{M^{2}_{4}}{2} \int d^4 x \, \sqrt{-g}\, R(g) + \int d^4 x \mathcal{L}_{M} (g,\psi)
    \end{equation}
where the $5$d bulk coordinates and the $5$d metric are respectively $X^A$ and $G_{AB}(X)$ with $A,B=0,1\dots,4$;
the $4$d brane coordinates and the $4$d metric are respectively $x^{\mu}$ and $g_{\mu \nu }(x)$ with $\mu,\nu=0,1\dots,3$ and the $\psi(x)$ are the $4$d matter fields which are described by the lagrangian $\mathcal{L}_{M}$ and are assumed to be confined to the $4d$ brane.
The constants $M_{5}$ and $M_{4}$ are respectively the $5$d and the $4$d Plank masses.
    %

 Taking the expansion around flat space $g_{\mu \nu} = \eta_{\mu \nu} + h_{\mu \nu}$ (with $\det h_{\mu \nu} = h$) and fixing the proper gauge, one can write the effective $4$d DGP action in the following way \cite{Hinterbichler:2011tt}:
\begin{equation} \label{eq:DGP-S}
    S = \int \,d^{4} x\, \frac{M_{4}^2}{4} \left[ 
    \frac{1}{2} h_{\mu \nu} \mathcal{E}^{\mu \nu, \alpha \beta} h_{\alpha \beta} - \frac{1}{2} m \left( 
    h_{\mu \nu} \triangle h^{\mu \nu } - h \triangle h
    \right)
    \right] + \frac{1}{2} h_{\mu \nu} T^{\mu \nu}
\end{equation}
where $\mathcal{E}^{\mu \nu, \alpha \beta}$ is the massless graviton kinetic operator defined by
\begin{equation}
    S_{\text{massless graviton}} = \int \, d^D x \, \frac{1}{2} h_{\mu \nu} \mathcal{E}^{\mu \nu, \alpha \beta} h_{\alpha \beta} 
\end{equation}
and 
\begin{equation}
    \mathcal{E}^{\mu \nu}_{\ \ \alpha \beta} = \left(
    \eta^{(\mu}_{\alpha} \eta^{\nu)}_{\beta} - \eta^{\mu \nu} \eta_{\alpha \beta} 
    \right) \Box  - 2 \partial^{(\mu} \partial_{(\alpha} \eta^{\nu)}_{\beta)} + \partial^{\mu} \partial^{\nu} \eta_{\alpha \beta} + \partial_{\alpha} \partial_{\beta} \eta^{\mu \nu}
\end{equation}
and, 
\begin{equation}
\label{scale_m}
    m \equiv \frac{2 M_{5}^3}{M_{4}^{2}}
\end{equation}
is the socalled DGP scale.
Interestingly the operator $\triangle$ is the (formal) square root of the $4$-dimensional Laplacian, i.e.,
\begin{equation} \label{op-DGP}
    \triangle \equiv \sqrt{- \Box}.
\end{equation}
Therefore, the action \eqref{eq:DGP-S} is of the Fierz-Pauli form with an operator dependent mass $m \triangle$. 
The operator \eqref{op-DGP} corresponds, in terms of our previous analysis, to the particular case $\alpha=1$ in \eqref{eq:nonlocTOloc}
and is known
as a resonance mass, or \emph{soft} mass~\cite{Hinterbichler:2011tt}. The particle content of the theory emerges clearly from a study of the propagator. By analiticity considerations (similar to those given in subsection \ref{subsec:unparticle}) the momentum part of the propagator can be reconstructed in terms of its poles and cuts and it is possible to write the following spectral representation:
\begin{subequations}
\label{spectral_representation_massive}
\begin{align}
\label{spectral_representation_massive_a}
\frac{-i}{p^2 +m\sqrt{p^2}}&= \int_0^\infty  ds\, \frac{-i}{p^2 +s}   \, \rho(s) \\
\rho(s)&= \frac{m}{\pi\sqrt{s}(s+m^2)} \, >0
\label{spectral_operator_mass}
\end{align}
\end{subequations}
 which shows therefore how  this theory
contains a continuum of ordinary (nonghost, nontachyon)
gravitons, with masses ($\sqrt{s}$) ranging from 0 to $\infty$. This is in line with what we would expect from the previous analysis using the extension problem for a nonlocal theory. The different form of the spectral function in Eq.~\eqref{spectral_operator_mass} relative to what we found in Eq.~(\ref{prop_nonlocal}) is due to the fact that, as opposed to our nonlocal  theory, Eq.~(\ref{final_non_local_action}), or equivalently  the unparticle model, Eq.~(\ref{unparticle_effective_action}), the DGP model is not scale invariant: it is indeed defined by the scale $m$ obtained as in Eq.~(\ref{scale_m}). The  behaviour in terms of a continuous spectrum of mass states is also understood as the result of the dimensional reduction of the noncompact fifth dimension, when the Kaluza-Klein tower collapses into a Kaluza-Klein continuum.
We conclude this paragraph with a remark about two limiting cases of the representation in Eq.~(\ref{spectral_representation_massive_a}). First consider the $m\to 0$ limit: the left hand side of Eq.~(\ref{spectral_representation_massive_a}) evidently reduces to the massless propagator $-i/p^2$ and the same is obtained by noting that in the same limit $\rho(s) \to 2\delta(s)$ and the factor of 2 is accounted for by the integration over half of the real axis. More interesting is perhaps the $m\to \infty$ limit
(or $m^2\gg p^2$) which is obtained by exchanging the limit with the integration in the right hand side of Eq.~(\ref{spectral_representation_massive_a}): \begin{equation}
\label{representation_limit}
    \frac{-i}{m \sqrt{p^2}} = \frac{1}{m} \int_0^\infty ds\, \frac{1}{\pi \sqrt{s}}\, \frac{-i}{p^2 +s}
\end{equation}%
Canceling out the mass scale $m$ in the above  Eq.~\eqref{representation_limit} one is left precisely with Eq.~\eqref{spectral_frac_propagator} upon replacing $m$ into $\sqrt{s}$ reproducing therefore the fractional propagator with $\alpha=1$. Thus the $m\to \infty$ limit in Eq.~(\ref{spectral_representation_massive_a}) amounts to going form a theory explicitly dependent on a scale $m$ over to a scale invariant theory ($m$ independent). The specific spectral density $1/(\pi\sqrt{s})$ is thus clearly connected to the scale invariance of the fractional propagator $-i/\sqrt{p^2}$.
This in retrospect means therefore that, for generic $\alpha$, the specific form of the function $\mu(\lambda)\propto \lambda^{1-\alpha}$ given in Eq.~(\ref{mufuntion}) in order to obtain the completeness relation, Eq.~(\ref{completeness}), is ultimately closely related to the request of scale invariance. Note that the representation in Eq.~(\ref{representation_limit}) coincides identically (up to a normalization constant) with Eq.~(\ref{prop_nonlocal}) for $\alpha=1$ and $\lambda^2\to s$ and a Wick rotation.

Another interesting connection with GR and cosmology can be obtained noting that the Proca Lagrangian that describes the three d.o.f of a massive photon 
\begin{equation}
    \mathcal{L}=-\frac{1}{4} F_{\mu \nu} F^{\mu \nu} - \frac{1}{2}m_{\gamma}^2 A_{\mu} A^{\mu}
\end{equation}
 has been shown to be equivalent to a gauge-invariant but nonlocal Lagrangian given by (see \cite{Belgacem:2017cqo})
\begin{equation}
    \mathcal{L}=-\frac{1}{4} F_{\mu \nu} \left( 1 - \frac{m_{\gamma}^2}{\Box} \right) F^{\mu \nu}
\end{equation}
therefore can be an interesting testbed for the strategy proposed in the previous section: rewrite a nonlocal action as a local action in $d+1$ dimensions using the extension problem, perform the calculations in this local setting and then read the result for the nonlocal action on the \emph{boundary}.  Moreover, it is a way to understand more difficult nonlocal invariant like $(\Box)^{-n}$. In the case of the Proca action we know that the final result should be a massive propagator.
The presence of nonlocal terms like inverse powers of the d'Alambertian can be interesting because these operators become relevant in the IR and therefore can have consequences in cosmology. Recent models are discussed 
\cite{Maggiore:2013mea,Belgacem:2018wtb, Deser:2019lmm, Amendola:2019fhc}.
In particular, the model proposed in \cite{Maggiore:2013mea} that considers nonlocal infrared corrections via the inverse Laplacian is promising for the cosmological implications while passing all the important tests at solar system scales \cite{Belgacem:2018wtb}.

\section{Conclusions}
In this paper, we have analyzed in detail the quantum aspects of a nonlocal fractional theory on the brane from the point of view of a (local)  scalar bulk quantum field theory with a boundary using the well-known and  and  well-defined mathematical framework of the ``extension problem'' developed by Caffarelli and Silvestre. 
In particular, we have focused on a nonlocal theory on the brane (for example a $(3+1)$-dimensional spacetime) defined in terms of the fractional Laplacian and studied it from the point of view of the local theory in the bulk --$(4+1)$-dimensional -- with one additional transverse space-like dimension ($y$). The problem is how to characterize the bulk fields that can describe the behavior of the nonlocal theory on the boundary. This is achieved by  solving  the equation of motion in the bulk separating out the transverse dimension ($y$), solving explicitly for the related eigenfunctions and imposing appropriate boundary conditions.
We have then  discussed the quantization of the local action in $(4+1)$ dimensions via the operator formalism though giving up local commutativity.  We finally computed the two-point bulk correlation functions that induce, on the brane ($y\to 0$ limit), the expected nonlocal correlations functions.

We find that what emerges naturally from the  standard quantization of the bulk theory via the canonical operator formalism is that the two-point Wightman function (or Feynman propagator) for the brane field can be expressed as an integral over the mass of the corresponding Wightman function (or Feynman propagator)  of an ordinary  massive scalar field  with a specific spectral density.   This allows us to make a connection with the scale invariant theory proposed by Georgi (unparticle model), showing that the two approaches are characterized by the same particle content and are therefore equivalent.

We have also discussed the brane vacuum energy of the nonlocal fractional theory  as obtained from the local Hamiltonian of the extended theory in  ($d$+1) dimensions. The bulk Hamiltonian operator ${\cal H}$ can be obtained as the component $T_{00}$ of the energy momentum tensor $T_{\mu\nu}$ of the $(d+1)$ local theory. Then, one can extract the vacuum brane energy from the bulk energy  ${E}_{\text{vac}}^{\text{brane}}$. This has been done, using an effective field theory approach therefore integrating out the degrees of freedom above a certain energy scale ($\Lambda_{\cal U}$) defined as the inverse of the  extra-dimension cutoff $y_\text{cut}$. This physical cutoff defines, on the brane, an effective field theory up to the energy $\Lambda_{\cal U}$.

We also extended the derivation of the eigenfunctions to the case of negative powers of the fractional Laplacian. The solution can be explicitly found in terms of the eigenfunctions of the initial problem changing the boundary conditions. Indeed, while the fractional Laplacian is an operator that maps Dirichlet boundary conditions to Neumann-type conditions for the local extended problem; the inverse fractional Laplacian is an operator that maps a Neumann-type boundary condition to a Dirichlet condition for the extended problem, at least in the restricted regime of validity $\alpha \in (0,2)$.

It is the authors' opinion that approaching the quantization of nonlocal fractional field theories via the extension problem should  be further explored (for instance with respect to vector and/or spinor fields) and exploited to study possible applications in general relativity and cosmology.

\subsection*{Acknowledgments}
The work of AMF was supported by a Swiss Government Excellence Scholarship
and is currently supported from ERC Advanced Grant GravBHs-692951 and MEC grant
FPA2016-76005-C2-2-P.\\
The authors wish to thank Michele Maggiore for many detailed and useful discussions.

\bibliographystyle{JHEP}
\bibliography{nonlocal} 
\end{document}